\documentclass[a4paper,12pt]{article}

\usepackage{amsmath}
\usepackage{amssymb}
\usepackage{amsthm}
\usepackage{mathrsfs}
\usepackage[T1]{fontenc}
\usepackage{enumerate}
\usepackage{fullpage}
\usepackage{color}

\def\mtr{matri}
\def\eqn{equation}
\def\cond{condition}

\def\tfn{transformation}

\def\fn{function}

\def\pl{Poisson--Lie}
\def\JL{Jacobi--Lie}

\def\JLtfn{Jacobi--Lie transformation}
\def\dd{Drinfel'd double}

\def\vbe{vanishing beta function equations}
\def\4diml{four-dimensional}
\def\bkg{background}
\def\wrt{with respect to}

\def\-1{^{-1}}
\def\half{\frac{1}{2}}

\def\e{{e}}

\def\wh{\widehat}

\def\PL{Poisson--Lie }
\def\pltp{Poisson--Lie T-pluralit}
\def\pltd{Poisson--Lie T-dualit}

\def\JLtp{Jacobi--Lie T-pluralit}

\def\gsugra{Generalized Supergravity Equation}
\def\usugra{usual Supergravity Equation}

\def\cf{{\mathcal {F}}}

\newcommand{\DDp}{DD$^+$}
\newcommand{\Exp}[1]{\operatorname{e}^{#1}}

\newcommand{\abs}[1]{\lvert {#1} \rvert}

\newcommand{\gLie}{\hat{\pounds}}

\newcommand{\unit}{\mathbf{1}}
\newcommand{\nul}{\mathbf{0}}

\newcommand{\D}{\mathscr{D}}
\newcommand{\G}{\mathscr{G}}

\newcommand{\cB}{\mathcal B}
\newcommand{\cD}{\mathcal D}
\newcommand{\cE}{\mathcal E}\newcommand{\cF}{\mathcal F}
\newcommand{\cG}{\mathcal G}
\newcommand{\cJ}{\mathcal J}

\newcommand{\cW}{\mathcal W}

\DeclareMathOperator{\arctanh}{Arctanh}

\begin{document}
\title{Jacobi--Lie Models and Supergravity Equations}
\author{Ladislav Hlavat\'y\footnote{ladislav.hlavaty@fjfi.cvut.cz}
\\ {\em Faculty of Nuclear Sciences and Physical Engineering,}
\\ {\em Czech Technical University in Prague,}
\\ {\em Czech Republic}
\and
Ivo Petr\footnote{ivo.petr@fit.cvut.cz}
\\ {\em Faculty of Information Technology,}
\\ {\em Czech Technical University in Prague,}
\\ {\em Czech Republic}}

\maketitle

\abstract{
Poisson--Lie T-duality/plurality was recently generalized to Jacobi--Lie T-plurality formulated in terms of Double Field Theory and based on Leibniz algebras given by structure coefficients $f_{ab}{}^{c},f_{c}{}^{ab},$ and  $Z_a,Z^a$. We investigate three- and four-dimensional sigma models corresponding to six-dimensional Leibniz algebras with  $f_b{}^{ba}\neq 0$, $Z^a=0$. We show that these algebras are plural one to another and, moreover, to an algebra with $f_b{}^{ba}= 0$, $Z^a=0$. These pluralities are used for construction of  Jacobi--Lie models. 

It was conjectured that plural models should satisfy \gsugra s. We have found examples of models satisfying ``true'' \gsugra s where no trivialization to \usugra s is possible. On the other hand, we show that there are also models corresponding to algebras with  $f_b{}^{ba}\neq 0$, $Z^a=0$ where the Killing vector appearing in \gsugra s either vanishes or can be removed by suitable gauge transformation. Such models then satisfy \usugra s, i.e. \vbe.
}

\tableofcontents


\section{Introduction}

Dualities play an important role in the study of string theory. Whereas Abelian T-duality \cite{buscher:ssbfe} (the T stands for target-space) is recognized as a symmetry of the theory, non-Abelian T-duality \cite{delaossa:1992vc} is rather understood as a solution-generating technique. In both cases we rely on the presence of symmetries of non-linear sigma-model characterizing dynamics of a string propagating in generally curved \bkg. These symmetries have to be gauged in order to obtain ``dual'' \bkg. With the introduction of \pl\ T-duality \cite{klise, klim:proc} and plurality \cite{unge:pltp} it became clear that isometries of sigma-model \bkg s are not strictly necessary and mutually dual models can be constructed on Lie groups $\G$ and $\G^*$ of \dd\ $\D$ with corresponding Lie bialgebra $(\bf g, \bf g^*)$. Such models are said to be \pl\ symmetric.

Revived interest in \pl\ T-duality appeared recently with its extension to U-duality \cite{saka:uext, malthom, btz} which is based on the exceptional Drinfel'd algebra (EDA) \cite{mst, hla:lalg}. While the algebra of the \dd\ is a Lie algebra, meaning that the structure constants satisfy $X_{AB}{}^C = - X_{BA}{}^C$, this doesn't have to hold for EDA.

An extension of Lie bialgebra $(\bf g, \bf g^*)$ was also introduced in \cite{rezaseph}, where the authors considered Jacobi structures on Lie groups \cite{iglemar} and Jacobi-Lie bialgebras $\left(\left({\bf g},\phi_0\right),\left({\bf g^*}, X_0\right)\right)$. Carefully repeating the process carried out in the derivation of \pltd y in \cite{klise} authors of \cite{rezaseph} found conditions for cocycles $\phi_0, X_0$ that allowed them to generalize \pl\ symmetry to \JL\ symmetry replacing Lie bialgebra by \JL\ bialgebra as the underlying algebraic structure. Low-dimensional \JL\ bialgebras and their Jacobi structures were further studied and classified in \cite{rezaseph:class, ahkr}.

It is natural to ask, whether models obtained this way satisfy \gsugra s 
\cite{Wulff:2016tju,sugra2} that for bosonic fields in the NS-NS sector read
\begin{align}\label{betaG}
0 &= R_{\mu\nu}-\frac{1}{4}H_{\mu\rho\sigma}H_{\nu}^{\
\rho\sigma}+\nabla_{\mu}X_{\nu}+\nabla_{\nu}X_{\mu},\\ \label{betaB}
0 &=
-\frac{1}{2}\nabla^{\rho}H_{\rho\mu\nu}+X^{\rho}H_{\rho\mu\nu}+\partial_{\mu}X_{\nu}-\partial_{\nu}X_{\mu},\\
\label{betaPhi} 0 &=
R-\frac{1}{12}H_{\rho\sigma\tau}H^{\rho\sigma\tau}+4\nabla_{\mu}X^{\mu}-4X_{\mu}X^{\mu}.
\end{align}
Here $R_{\mu\nu}$ and $R$ are Ricci tensor and scalar curvature of metric $\cG$ specifying sigma-model \bkg\ $\cf= \cG+\cB$ together with Kalb-Ramond field $\cB$ giving rise to torsion
\begin{equation}
 H_{\rho\mu\nu}=\partial_\rho {{\cB}}_{\mu\nu}+\partial_\mu {{\cB}}_{\nu\rho}+ \partial_\nu \cB_{\rho\mu}.
\end{equation}
Components of one-form $X$ are calculated from dilaton $\Phi$ and Killing vector field $\cJ$ as
\begin{equation}\label{xform}
X_\mu:=\partial_\mu\Phi+\cJ^\kappa\cf_{\kappa\mu}.
\end{equation}
The condition that the field $\cJ$ is Killing vector field of the \bkg\ $\cf$ is not necessary for solution of eqns. \eqref{betaG}-\eqref{betaPhi}, but it is required for their full version containing the R-R fields \cite{sugra2}. 

Note that $X$ is  invariant \wrt\ gauge \tfn\
\begin{equation}\label{gauge tfn lambda}
\Phi_\Lambda:=\Phi +\Lambda, \quad \cJ_\Lambda:= \cJ-\mathrm{d}\Lambda \cdot \cf\-1
\end{equation}
where $\Lambda$ is arbitrary differentiable function. It means that if $\mathrm{d}X=0$,  we can always find dilaton $\Phi_\Lambda$ such that $X_\mu=\partial_\mu\Phi_\Lambda$, $\cJ_\Lambda$ vanishes, and the \eqn s \eqref{betaG}-\eqref{betaPhi} transform to \usugra s, i.e. 
\vbe. Moreover, it immediately follows from the \eqn\  \eqref{betaB} that $\mathrm{d}X=0$ whenever torsion $H_{\rho\mu\nu}$ vanishes. 
If $\mathrm{d}X\neq 0$, we cannot trivialize  Killing vector field $\cJ$ by gauge \tfn\  and the \eqn s \eqref{betaG}-\eqref{betaPhi} are called \gsugra s.
More general conditions for trivialization of the Killing vector were considered in \cite{wulff}.

Authors of \cite{melsaka} approach \JLtfn\ from the perspective of Double Field Theory (DFT) \cite{hullzw} that has proven useful earlier in the study of \pltd y \cite{hass, dehath, saka2}. Similarly to EDA they break the antisymmetry of structure constants $X_{AB}{}^C$ and propose $2D$-dimensional Leibniz algebra DD$^+$ composed of $D$-dimensional subalgebras $\bf g$ (generated by $T_a$) and $\bf g^*$ (generated by $T^a$) in the form
\begin{align}
\begin{split}
 T_{a}\circ T_{b} &= f_{ab}{}^c\,T_{c}\,, \qquad
 T^a\circ T^b = f_c{}^{ab}\,T^c \,,
\\
 T_{a}\circ T^b &= \bigl(f_a{}^{bc} + 2\,\delta_a^b\,Z^c -2\,\delta_a^c\,Z^b\bigr)\,T_c - f_{ac}{}^b\,T^c +2\,Z_a\,T^b\,,
\\
 T^a\circ T_b &= - f_b{}^{ac}\,T_{c} +2\,Z^a\,T_b + \bigl(f_{bc}{}^a +2\,\delta^a_b\,Z_c -2\,\delta^a_c\,Z_b\bigr)\,T^c\,,
\end{split}
\label{eq:DD+components}
\end{align}
where $a=1,\dotsc,D$. {Since $\bf g$ and $\bf g^*$ are Lie algebras, antisymmetry holds for their structure constants $f_{ab}{}^c =-f_{ba}{}^c$ and $f_c{}^{ab}=-f_c{}^{ba}$. A symmetric bilinear form can be introduced on DD$^+$, such that the subalgebras $\bf g$ and $\bf g^*$ are maximally isotropic with respect to it, i.e. it holds that
\begin{align}
 \langle T_a,\,T^b \rangle = \delta_a^b\,,\qquad
 \langle T_a,\,T_b \rangle = \langle T^a,\,T^b \rangle = 0.\label{excmat}
\end{align}
As opposed to the case of Lie bialgebra, the form $\langle , \rangle$ on DD$^+$ is no longer ad-invariant.} Under certain conditions on $f_{ab}{}^c, f_c{}^{ab}, Z_a$ and $Z^a$ following from Leibniz identities there is one-to-one correspondence between Leibniz algebra DD$^+$ of \cite{melsaka} and Jacobi-Lie bialgebra of \cite{rezaseph}. In the following we prefer the approach based on DD$^+$ and DFT since it not only allows us to investigate \JLtfn\ of \bkg\ fields $\cG$ and $\cB$, but also enables us to study \tfn\ of dilaton and vector field $\cJ$ that appear in generalized fluxes of DFT. Moreover, the construction of DFT fields and corresponding sigma-model \bkg s is not limited to coboundary \JL\ bialgebras.

When performing \pl\ T-duality \tfn\ with respect to a group whose structure constants are non-unimodular, $f_b{}^{ba} \neq 0$, one finds that resulting background does not always satisfy \vbe\ \cite{grv,aagl} but \gsugra s \eqref{betaG}-\eqref{betaPhi} instead \cite{hkc}. In such cases the transformed dilaton may depend on dual coordinates that have to be eliminated at cost of introducing Killing vector field $\cJ$ \cite{saka2, HP}. 

In \JL\ symmetric models the presence of non-vanishing $Z^a$ in DD$^+$ gives rise to a scale factor that brings the dependence on dual coordinates even in the metric and $\cB$-field. Even if this is consistent from the point of view of DFT, we can hardly consider such fields as solutions to (Generalized) Supergravity Equations.

There are three types of models considered in Section 3 of \cite{melsaka} depending on values of $f_b{}^{ba}$ and $Z^a$. Examples with $f_b{}^{ba}=Z^a=0$ belong to the first case and satisfy \usugra s, i.e. \eqref{betaG}-\eqref{betaPhi} with vanishing $\cJ$. Examples with $Z^a \neq 0$ belong to the third case and give \bkg s depending on dual coordinates. In this paper we want to investigate models of type 2, i.e. those obtained from Leibniz algebras with $f_b{}^{ba}\neq 0$ and $Z^a=0$. Our primary goal is to find solutions of \gsugra s predicted in \cite{melsaka}. We discuss construction of $\cJ$ and give examples where \gsugra s reduce to \usugra s  even in this case. 

In the language of DFT it is natural to consider not only duality, but also plurality. When viewing DD$^+$ as a vector space, plurality allows us to pass to different pair of subalgebras $\bf g$ and $\bf g^*$ instead of just exchanging them. We shall show that three-dimensional Jacobi-Lie bialgebras with $f_b{}^{ba}\neq 0, Z^a=0$ classified in Table 7 of \cite{rezaseph:class} are isomorphic to type 1 algebra $\left(\left({\bf g},\phi_0\right),\left({\bf g^*}, X_0\right)\right)={\left(({III},-T^2+T^3),({I},0)\right)}$ and use this plurality to construct our examples.

Plan of the paper is the following. In Section \ref{sec:algebras} we summarize notation used to describe the algebra \DDp\ and results concerning plurality of \JL\ bialgebras. The particular isomorphisms are given in the Appendix. In Section \ref{sec:JLmodels} we recapitulate the construction of \JL\ sigma model \bkg s and find sigma model satisfying the Supergravity Equations. Section \ref{sec:pltp} sums up \JLtp y and appearance of vector field $\cJ$. Examples of plural \bkg s without spectators are presented in Section \ref{sec:plural_models} and examples of \JLtp y with spectators are in Section \ref{sec:spec}.

\section{Leibniz algebras with  $f_b{}^{ba}\neq 0, Z^a=0$} \label{sec:algebras}

The $2D$-dimensional Leibniz algebra DD$^+$ was introduced in \cite{melsaka}. Denoting $T_A = (T_a, T^a)$, $Z_A = (Z_a, Z^a)$,  $a = 1,\ldots,D$; $A = 1,\ldots,2D$, the structure constants of DD$^+$ can be written as
\begin{equation}
T_A \circ T_B = X_{AB}{}^C \, T_C, \label{dd_st_const}
\end{equation}
while equations \eqref{excmat} read
\begin{equation}\label{eta}
\langle T_A, T_B \rangle = \eta_{AB}, 
\qquad 
\eta_{AB} = \begin{pmatrix} 0 & \delta_a^b \\ \delta_a^b & 0 \end{pmatrix}.
\end{equation}

There is one-to-one correspondence between Leibniz algebras \eqref{eq:DD+components} and Jacobi-Lie bialgebras studied in \cite{rezaseph,rezaseph:class} and we can use the classification of three-dimensional Jacobi-Lie bialgebras given in \cite{rezaseph:class}. Investigating Table 7 of \cite{rezaseph:class} one can find that all algebras with  $f_b{}^{ba} \neq  0$ and $Z^a=0$ appear as duals of the last seven bialgebras. We give their list\footnote{The first two columns of Table \ref{tab:algebras} refer to Bianchi classification of three-dimensional Lie algebras. The commutation relations and cocycles of \cite{rezaseph,rezaseph:class} are restored by setting $[T_A, T_B]:=\frac{1}{2}(T_A \circ T_B - T_B \circ T_A)$ and $\phi_0 = \beta_a T^a := 2 Z_a T^a$, $X_0 = \alpha^a T_a := 2 Z^a T_a$.} in Table \ref{tab:algebras}, where algebra $\bf g^*$ is always defined as
\begin{equation}
T^1 \circ T^2 = -(T^2 + T^3), \quad T^1 \circ T^3 = -(T^2 + T^3).
\end{equation}

\begin{table}
\begin{center}
\begin{tabular}{| l | l | l |l |}
\hline
{\footnotesize ${\bf g}$ }& {\footnotesize ${\bf g^{*}}$}
&{\footnotesize Product definitions of ${\bf g}$}&{$\phi_{0}=2 Z_a T^a$}\\
\hline
\hline
{\footnotesize $III.v$}&{\footnotesize $III$}&{\footnotesize $T_1\circ T_3=T_1,T_2\circ T_3=T_1$}&{\footnotesize $\frac{1}{2}({T}^{2}-{T}^{3}$})\\
\hline
{\footnotesize $III.x$}&{\footnotesize $III$}&{\footnotesize $T_1\circ T_2=T_1,T_1\circ T_3=-T_1$}&{\footnotesize $-({T}^{2}-{T}^{3})$}\\
\hline
{\footnotesize $IV.iv$}&{\footnotesize $III$}&{\footnotesize $T_1\circ T_2=-T_1,T_1\circ T_3=T_1,T_2\circ T_3=T_1+T_2+T_3$}&{\footnotesize ${T}^{2}-{T}^{3}$}\\
\hline
{\footnotesize $V.iii$}&{\footnotesize $III$}&{\footnotesize $T_1\circ T_2=-T_1,T_1\circ T_3=T_1,T_2\circ T_3=T_2+T_3$}&{\footnotesize ${T}^{2}-{T}^{3}$}\\
\hline
{\footnotesize $VI_{0}.iv$}&{\footnotesize $III$}&{\footnotesize $T_1\circ T_2=-T_1,T_1\circ T_3=T_1,T_2\circ T_3=-(T_2+T_3)$}&{\footnotesize ${T}^{2}-{T}^{3}$}\\
\hline
{\footnotesize $VI_{a\pm}$}&{\footnotesize $III$}&{\footnotesize $T_1\circ T_2=T_1,T_1\circ T_3=-T_1,T_2\circ T_3=-a_\pm(T_2+T_3)$}&{\footnotesize $-({T}^{2}-{T}^{3})$}\\
& & {\footnotesize $a_\pm=\frac{a\pm1}{a\mp1},a\neq1$} & \\
\hline
\end{tabular}
\caption{{Real three-dimensional type 2 Leibniz algebras with $X_{0}=2 Z^a\,T_a=0$, $f_b{}^{ba}\neq 0$.} \label{tab:algebras}}
\end{center}
\end{table}

Maybe surprisingly, \emph{all algebras in Table \ref{tab:algebras} are isomorphic} in the sense that there are matrices $C$ that transform algebraic relations of algebra generated by $T_A$ to those of algebra generated by $\hat T_A$ as
\begin{equation}\label{Cplurality}
\hat T_A = C_A{}^B \, T_B , \quad \hat T_A\circ \hat T_B=\hat X_{AB}{}^C \,\hat T_C.
\end{equation}
Since \eqref{eta} has to hold as well, the conditions on $C$ are
\begin{equation} \label{tfnX}
C_{A}{}^{F}C_{B}{}^{G} X_{FG}{}^H = \hat X_{AB}{}^D\,C_{D}{}^{H}, \qquad C_{A}{}^{F}C_{B}{}^{G} \eta_{FG}=\eta_{AB}.
\end{equation}
Moreover, all algebras in Table \ref{tab:algebras} are also isomorphic to algebra $\left(({\bf g},\phi_0),({\bf g^*}, X_0)\right)=\left((III,-T^2+T^3),(I,0)\right)$ defined by 
\begin{equation}\label{com_rel:type1}
T_1\circ T_2=-(T_2+T_3), \quad T_1\circ T_3=-(T_2+T_3), \quad T^a\circ T^b=0,
\end{equation}
and 
\begin{equation}\label{cocycles:type1}
\phi_0 =2 Z_a T^a = -T^2+T^3,\qquad X_{0}=2 Z^a\,T_a=0.
\end{equation}

For better readability and similarity to the Manin triples we shall denote the Leibniz algebras using curly brackets as
\begin{align*}
{\left(\left({III},-T^2+T^3 \right),\left({I},0\right)\right)}&=:\{3|1\},\\
{\left(\left({III.v},\half(T^2-T^3)\right),\left({III},0\right)\right)}&=:\{3v|3\},\\
{\left(\left({III.x},-T^2+T^3 \right),\left({III},0\right)\right)}&=:\{3x|3\},\\
{\left(\left({IV.iv},T^2-T^3\right),\left({III},0\right)\right)}&=:\{4iv.3\},\\
{\left(\left({V.iii},T^2-T^3\right),\left({III},0\right)\right)}&=:\{5iii|3\},\\
{\left(\left({VI_0.iv},T^2-T^3\right),\left({III},0\right)\right)}&=:\{6_0iv|3\},\\
{\left(\left({VI_{a\pm}},-T^2+T^3 \right),\left({III},0\right)\right)}&=:\{6_{a\pm}|3\}.
\end{align*}
The existence of isomorphisms then can be written as 
$$ \{3|1\}\cong \{3v|3\}\cong \{3x|3\}\cong \{4iv|3\}\cong \{5iii|3\}\cong \{6_0iv|3\}\cong \{6_{a\pm}|3\}.$$
In each case there are several possible solutions to algebraic relations \eqref{tfnX} that can be used for \JL\ plurality \tfn\ and different solutions may give models with different properties. The \tfn\ matrices $C$ are given in the Appendix. 

Our strategy for getting \JL\ models satisfying \gsugra s is following. We start with flat torsionless model with vanishing dilaton on the group corresponding to the algebra with $f_b{}^{ba}=Z^a=0$ and then use \JLtp y to get models with $f_b{}^{ba}\neq 0$ and $Z^a=0$.

\section{\JL\ models}\label{sec:JLmodels}

The construction of \JL\ symmetric models resembles construction of \PL symmetric models \cite{melsaka}. In the absence of spectator fields the \bkg\ fields are given by constant matrix $E_0$ and algebraic structure of Leibniz algebra DD$^+$. The metric $\cG$ and the $\cB$-field of \JL\ model on Lie group $\G$ corresponding to Lie algebra ${\bf g}$ can be expressed as 
\begin{align}\label{JLmtz}
 \cF_{mn} \equiv \cG_{mn}+\cB_{mn} = \Exp{-2\omega} R_{ab}\,r_m^a\,r_n^b ,\quad
 (R_{ab}) \equiv ((E_0{}\-1){}^{ab}+\pi^{ab})^{-1}\,
\end{align}
where $r_m^a$ are components of right-invariant one-form $\mathrm{d}g g\-1=r_m^a \mathrm{d}x^m T_a$ and $g \in \G$. The anti-symmetric matrix $\pi^{ab}$ is calculated as follows. For group action
\begin{equation*}
g \triangleright T_A := \Exp{x^b T_b \circ} T_A
\end{equation*}
matrix $M_A{}^B$ defined as
\begin{equation}
g\-1 \triangleright T_A := M_A{}^B T_B
\end{equation}
is of the form
\begin{equation}\label{matMAB}
M_A{}^B = 
\begin{pmatrix} 
a_a^b & 0 \\
-\pi^{ac}\,a_c^b & \Exp{-2\Delta} (a\-1)^b_a
\end{pmatrix},
\end{equation}
from which we can read $\pi^{ab}, \Delta$ and define \JL\ structure
\begin{equation}\label{piDelta}
\pi^{mn} = \Exp{2\Delta} \pi^{ab} e_a^m e_b^n, \qquad r_m^a e_b^m = \delta_b^a
\end{equation}
on $\G$. 

For the factor $\Exp{-2\omega}$ we have to use DFT. Let us denote the doubled coordinates $x^M = (x^m,\tilde x_m)$ and corresponding derivatives $\partial_M = (\partial_m,\tilde \partial^m)$. The generalized frame fields
\begin{equation}
E_A{}^M = \begin{pmatrix} e_a^m & 0 \\ -\pi^{ab}\,e_b^m & \Exp{-2\omega}r^a_m  \end{pmatrix}, \qquad \Exp{-2\omega} := \Exp{-2\Delta} \tilde{\sigma}
\end{equation}
satisfy condition for generalized Lie derivative $\gLie$ in DFT
\begin{align}
 \gLie_{E_A} E_B{}^M = -X_{AB}{}^C\,E_C{}^M
\label{eq:JL-frame-algebra}
\end{align}
if conditions
\begin{equation}
\partial_m \tilde{\sigma} = 0, \qquad \tilde \partial^m \tilde{\sigma} = -2 Z^a v_a^m
\end{equation}
hold, where $v_a^m$ are components of left-invariant fields on $\G$. If $Z^a = 0$, we can choose $\tilde\sigma = 1$ and $\omega=\Delta$. For $Z^a \neq 0$ the \bkg s in \eqref{JLmtz} may be dependent on ``dual'' coordinates $\tilde x_m$.

The standard dilaton $\Phi$ can be found as
\begin{align}
 \Exp{-2\Phi} =\Exp{-2\varphi} \Exp{(D-1)\,\Delta} \tilde{\sigma}^{\frac{D-3}{2}} \,\abs{\det ((E_0{}\-1){}^{ab}+\pi^{ab}) \det(a_a{}^b)}
\label{eq:dilaton-JL}
\end{align}
Knowledge of \fn\ $\varphi$ is complementary to knowledge of dilaton. $\cG$, $\cB$ and $\Phi$ constructed above define sigma model on $\G$ and should satisfy equations \eqref{betaG}-\eqref{betaPhi}.

\subsection{Three-dimensional flat models corresponding to the algebra 
\normalsize{$\{3|1\}$}}\label{III.I}

As mentioned in the previous section, for getting models satisfying \gsugra s we are going to look for models corresponding to the algebra with $f_b{}^{ba}=Z^a=0$ and then apply the \JLtp y to construct models with  $f_b{}^{ba}\neq 0$, $Z^a=0$.

The algebra denoted in \cite{rezaseph:class} as $\left(({III},-T^2+T^3), ({I},0)\right)=:\{3|1\}$ is given by relations \eqref{com_rel:type1} and \eqref{cocycles:type1}. It is of the type 1, so the models should satisfy \usugra s. We parametrize elements $g\in \G$ as
$$ g= e^{x^1 T_1} e^{x^2 T_2} e^{x^3 T_3}.$$
Given $E_0$, it is then easy to calculate $\cF$. More difficult is to find dilaton that solves Supergravity Equations. However, one can proceed reversely and look for matrices $E_0$ that together with the algebra $\{3|1\}$ produce sigma model satisfying \usugra s with vanishing (or constant) dilaton $\Phi$. This will enable us to find the function $\varphi(x)$ in \eqref{eq:dilaton-JL} necessary for \JLtp y \tfn\ of dilaton. 

For vanishing $\Phi$ and $\cJ$ equation \eqref{betaB} implies $H_{\mu\nu\rho}=0$. Then from \eqref{betaG} and \eqref{betaPhi} one can see that the \JL\ \bkg\ has to be Ricci flat. In three dimensions that means that the metric is also flat. Altogether there are two families of matrices $E_0$ depending on real constants $\lambda_i$ that produce flat torsionless \bkg s via \eqref{JLmtz}, namely
\begin{equation}\label{E0diag}
E_0=\left(
\begin{array}{ccc}
 \frac{5 \lambda_3^2+2 \lambda_3 (-4 \lambda_4+8 \lambda_6+\lambda_7)+4 \lambda_4^2-16 \lambda_4 \lambda_6-16 \lambda_5^2+16 \lambda_6^2+\lambda_7^2}{ \lambda_3-\lambda_4+2 (\lambda_5+\lambda_6)} & \lambda_3-\lambda_4+\lambda_7 & \lambda_3 \\
 \lambda_4 & \lambda_5 & \lambda_6 \\
 \lambda_7 & \lambda_3-\lambda_4+\lambda_6 & \lambda_5 \\
\end{array}
\right)
,\end{equation}
and
\begin{equation}\label{E06}
E_0=\left(
\begin{array}{ccc}
 2 (\lambda_3-\lambda_4+\lambda_6-\lambda_8) & -\lambda_3-2 (\lambda_5+\lambda_6) & \lambda_3 \\
 \lambda_4 & \lambda_5 & \lambda_6 \\
 -\lambda_4-2 (\lambda_5+\lambda_8) & \lambda_8 & -\lambda_5-\lambda_6-\lambda_8 \\
\end{array}
\right).
\end{equation}
For simplicity we choose $\lambda_3 = \lambda_4 = \lambda_6 = \lambda_7 = 0$ and $\lambda_5=1$ in \eqref{E0diag} to work with diagonal $E_0$ and get \bkg\ of \JL\ sigma model\footnote{All background fields in the examples depend on coordinates $x^m$. For better readability we use subscripts $x_m$ to display coordinate indices.}
\begin{align}\label{flatmtz1}
\cF=e^{-2 x_1+x_2-x_3}
\left(
\begin{array}{ccc}
 -8 e^{2 x_1} & 0 & 0 \\
 0 & \cosh (2 x_1) & -\sinh (2 x_1) \\
 0 & -\sinh (2 x_1) & \cosh (2 x_1) \\
\end{array}
\right).
\end{align}
Choosing $\lambda_3=1, \lambda_4=0, \lambda_6=\lambda_8=\frac{\lambda_5}{3}=-\frac{1}{8}$ the $E_0$ in \eqref{E06} becomes symmetric and \JL\ model is given by \bkg
\begin{equation}\label{flatmtz2}
\cF=e^{-2 x_1+x_2-x_3}\left(
\begin{array}{ccc}
 2e^{2 x_1} & -\frac{1}{2} \left(e^{2 x_1}-1\right) & \frac{1}{2} \left(e^{2 x_1}+1\right) \\
 -\frac{1}{2} \left(e^{2 x_1}-1\right) & \frac{1}{8}\left(e^{2 x_1}-4\right) & -\frac{1}{8}e^{2 x_1} \\
 \frac{1}{2} \left(e^{2 x_1}+1\right) & -\frac{1}{8}e^{2 x_1} & \frac{1}{8} \left(e^{2 x_1}+4\right) \\
\end{array}
\right).
\end{equation}
In both cases we get constant dilaton $\Phi$ from the expression \eqref{eq:dilaton-JL} e.g. for
\begin{align}\label{flatfi}
\varphi=x_1 -\frac{x_2}{2} + \frac{x_3}{2}.
\end{align}
Even though these two flat models must be equivalent by transformation of coordinates, they give different results after the \JL{}  \tfn\footnote{Similar situation appeared in pluralization of Minkowski metric \cite{HPP}.}.

Let us note that the structure of algebra $\{3|1\}$ leads to vanishing $\pi^{ab}$ in \eqref{matMAB}. We choose $\tilde\sigma = 1$ to have $\omega = \Delta = -\frac{x_2}{2} + \frac{x_3}{2}$ and get the overall factor $\Exp{x_2-x_3}$ in \eqref{flatmtz1} and \eqref{flatmtz2}. This is similar to construction of \PL symmetric \bkg s (where $Z_a=0$) on semi-Abelian Lie bialgebra $(3|1)$ where the Poisson bivector also vanishes. However, then we have $\Delta=0$, $\Exp{-2\Delta} = 1$, and we get \bkg s with constant scalar curvature (the sign of $R$ depends of the choice of $\lambda_i$ in $E_0$) that do not satisfy \gsugra s for any $\Phi$ or $\cJ$.

\section{\JLtp y}\label{sec:pltp}

\JLtp y was described in \cite{melsaka} in the DFT language. Transformation of metric and $\cB$-field is very similar to the \pltp y of models constructed from \dd s. Namely, we first transform the constant matrix $E_0$ by formula
\begin{equation}\label{E0hat}
\hat E_0=\left(\left(P+ E_0 \cdot R \right)^{-1} \cdot \left(Q+E_0 \cdot S \right)\right)^T
\end{equation}
where the matrices $P,Q,R,S$ are $D\times D$ blocks of the \tfn\ \mtr x $C^{-1}$
$$C^{-1}=\begin{pmatrix}
 P & Q \\
 R & S
\end{pmatrix}, \qquad T_A = (C^{-1})_A{}^B \, \hat T_B $$
and then apply formulas for construction of \JL\ models given in Section \ref{sec:JLmodels}.

More difficult is the \tfn\ of dilaton. Before we can apply the formula \eqref{eq:dilaton-JL} we have to transform function $\varphi$. For that we have to compute generalized fluxes $\cF_A$ associated with generalized vielbein $\cE_{A}{}^{M} \equiv \Exp{\omega(x)} E_A{}^M(x)$ and DFT dilaton $d$ through 
\begin{align}\label{FM}
\cF_A \equiv \cW^B{}_{AB} + 2\, \cD_A d  = \cE_A{}^M\, F_M=\Exp{\omega(x)}F_A
\end{align}
where
\begin{equation}
\cW_{ABC} \equiv - \cD_A \cE_B{}^M\, \cE_{MC} \,,\quad 
\cD_A \equiv \cE_A{}^M\,\partial_M,\quad
\Exp{-2d}=\Exp{-2\Phi}\sqrt{|\det \cG|} \,.
\end{equation}
Finally we find for $f_b{}^{ba}=0,\ Z^a=0$,
\begin{align}\label{gradfi}
 \partial_M \varphi= \half F_M .
\end{align}
Transformed function $\hat\varphi$ is then obtained from the \tfn\ of the flux $F_A:=\Exp{-\omega(x)}\cF_A$ 
\begin{equation}\label{hatFA}
\hat F_A=C_A{}^B F_B.
\end{equation}

However, for the algebras of type 2, where $\hat f_b{}^{ba}\neq0,\ \hat Z^a=0$, it may happen that  the r.h.s. of \eqref{gradfi} is not  gradient of a \fn\ of $x^m$. Instead, we may define form $\hat Y=\hat Y_M\,d\hat x^M$,
\begin{align}\label{rhs}
\hat Y_M:= \half \hat F_M  + \begin{pmatrix} \nul_D \\ \frac{ \hat f_b{}^{ba} \,\hat v_a^m}{2\tilde{\sigma}}\end{pmatrix}^T=
 \begin{pmatrix} \partial_m\hat \varphi(x) \\  \hat \cJ^m(x)
\end{pmatrix}^T
\end{align}
where $\hat v_a^m$ are components of left-invariant vector fields of the group associated with $\hat{\bf{g}}$. It turns out that $\hat \cJ^m(x)$ then are components of Killing vector field that occurs in the \gsugra s for the \JL\ plural model. Values of \fn s $\hat\varphi$ and $\hat \cJ^m$ for various type 2 algebras are given in the Table \ref{tab:killings s}. It is evident that if $\hat \cJ^m$ vanishes then corresponding plurality produces \JL\ model that satisfies \usugra s. On the other hand, there are cases with non-vanishing $\hat \cJ^m$ where $\mathrm{d}\hat X=0$ and we can get rid of the Killing vector field by the gauge \tfn\ \eqref{gauge tfn lambda}. The model then satisfies \usugra s - see e.g. \ref{3to3vflatmtz2}.  

Let us note in the end that Jacobi--Lie T-duality $T_a\leftrightarrow T^a$ for $Z_a\neq 0$ is not useful for construction of new models  as it transforms type 1 and type 2 algebras to the type 3 where $ \hat X_0\neq 0$, and these produce \JL\ models dependent on dual DFT coordinates.

\section{\JL\ models plural to $\{3|1\}$}\label{sec:plural_models}
When we have \tfn\ matrices $C$ we can construct supergravity fields  of \JL\ models plural to flat \bkg s found in \ref{III.I}. However, not all matrices $C$ generate plural models. Some of them may give singular matrix $\hat E_0$. Beside that it turns out that even for the type 2 algebras some \JL\ plural models satisfy usual and others \gsugra s. Models satisfying \gsugra s are produced only by \JL\ \tfn\ of the \bkg\ \eqref{flatmtz1}.  Unfortunately the \bkg s are usually so extensive that they are  difficult to display.

\begin{table}
\begin{center}
\begin{tabular}{|c | c || c | c |}
\hline
{\footnotesize Plural algebra  }& {\footnotesize C-matrix}
&{\footnotesize \fn\ $\hat\varphi$}&{\footnotesize Killing vector}\\
\hline
\hline
{\footnotesize$\{3.v|3\}$}&{\footnotesize $C_\alpha$}&{\footnotesize $-\half x_3$}&{\footnotesize  $(0,0,0)$ }\\
{\footnotesize $\{3.v|3\}$}&{\footnotesize $C_\beta$}&{\footnotesize $\frac{1}{4}(x_2+x_3)$}&{\footnotesize $(2\,e^{-x_3},0,0)$ }\\
{\footnotesize $\{3.v|3\}$}&{\footnotesize $C_\gamma$}&{\footnotesize $\frac{1}{4}(x_2+x_3)$}&{\footnotesize  $(2\,e^{-x_3},0,0)$ }\\
\hline
{\footnotesize $\{3.x|3\}$}&{\footnotesize $C_\alpha$}&{\footnotesize 0}&{\footnotesize  $(2\,e^{-x_2+x_3},0,0)$} \\
{\footnotesize $\{3.x|3\}$}&{\footnotesize $C_\beta$}&{\footnotesize $\frac{1}{2}(x_3-x_2)$}&{\footnotesize $(0,0,0)$ }\\
\hline
{\footnotesize $\{4.iv|3\}$}&{\footnotesize $C_\alpha$}&{\footnotesize $\frac{1}{2}(x_2-x_3)$}&$(2\,e^{x_2-x_3},0,0)$ \\
{\footnotesize $\{4.iv|3\}$}&{\footnotesize $C_\beta$}&{\footnotesize $0$}&{\footnotesize $(0,0,0)$ }\\
\hline
{\footnotesize $\{5.iii|3\}$}&{\footnotesize $C_\alpha$}&{\footnotesize $\frac{1}{2}(x_2-x_3)$}&{\footnotesize  $(2\,e^{x_2-x_3},0,0)$} \\
{\footnotesize $\{5.iii|3\}$}&{\footnotesize $C_\beta$}&{\footnotesize $0$}&{\footnotesize $(0,0,0)$}\\
\hline
{\footnotesize $\{6_0.iv|3\}$}&{\footnotesize $C_\alpha$}&{\footnotesize $x_2-x_3$}&{\footnotesize $(0,0,0)$} \\
{\footnotesize $\{6_0.iv|3\}$}&{\footnotesize $C_\beta$}&{\footnotesize $\frac{1}{2}(x_3-x_2)$}&{\footnotesize  $(2\,e^{x_2-x_3},0,0)$}\\
{\footnotesize $\{6_0.iv|3\}$}&{\footnotesize $C_\gamma$}&{\footnotesize $\frac{1}{2}(x_3-x_2)$}&{\footnotesize  $(2\,e^{x_2-x_3},0,0)$ }\\
\hline
{\footnotesize $\{6_{a\pm}.iv|3\}$}&{\footnotesize $C_\alpha$}&{\footnotesize  $\frac{a_{\pm}}{2}(x_3-x_2)$}&{\footnotesize $(2\,e^{-x_2+x_3},0,0)$}\\
{\footnotesize $\{6_{a\pm}.iv|3\}$}&{\footnotesize $C_\beta$}&{\footnotesize $\frac{a_\pm-1}{2}(x_2-x_3)$}&{\footnotesize $(0,0,0)$ }\\
\hline
\end{tabular}
\caption{Values of \fn s $\varphi$ and components of Killing vector fields for type 2 algebras and corresponding C-matrices from Appendix.}
\label{tab:killings s}
\end{center}
\end{table}

\subsection{Models obtained by \JLtp y of the  \bkg\ \eqref{flatmtz1}}

Below we present examples obtained by \JL\ \tfn\  of the flat model given by $\{3|1\}=\left(\left(III,-T^2+T^3\right),\left(I,0\right)\right)$ and 
\begin{equation}\label{E0diag1}
E_0=\left(
\begin{array}{ccc}
 -8 & 0 & 0 \\
 0 & 1 & 0 \\
 0 & 0 & 1 \\
\end{array}
\right).
\end{equation}
In detail we show results of \JLtp y only for the algebras $\{3.v|3\}$ and $ \{3.x|3\}.$ Analogous results appear for \JLtp ies to the other algebras in the Table \ref{tab:algebras}, namely $\{4.iv|3\}$, $\{5.iii|3\}$ and $\{6_0.iv|3\}$.  

\subsubsection{\JL\ models corresponding to $\{3.v|3\}$}\label{III. v}

There are three \mtr ces (see Appendix) that up to choice of parameters solve \cond s \eqref{tfnX} for isomorphism 
$$
\{3|1\} \cong \{3.v|3\}.
$$
Plural models exist only for 
\begin{equation}\label{cbeta3v}
C_\beta=\left(
\begin{array}{cccccc}
 0 & \frac{1}{4} & \frac{1}{4} & -1 & \frac{1}{2} &
   -\frac{1}{2} \\
 0 & 0 & \frac{1}{2} & 0 & 1 & 0 \\
 \frac{1}{2} & 0 & -\frac{1}{2} & 0 & 1 & 0 \\
 1 & 0 & 0 & 0 & 8 & 0 \\
 0 & -\frac{1}{2} & -\frac{1}{2} & 4 & -3 & 3 \\
 0 & -\frac{1}{2} & -\frac{1}{2} & 4 & -1 & 1 \\
\end{array}
\right)
\end{equation}
because matrix $({P}+ E_0 \cdot {R})$ in the formula \eqref{E0hat} is singular for $C_\alpha,C_\gamma$ and the matrices  $\hat E_0$ do not exist in these cases.

For \eqref{cbeta3v} we get
$$\hat E_0=\left(
\begin{array}{ccc}
 \frac{1}{24} & -\frac{1}{12} & -\frac{1}{12} \\
 \frac{1}{4} & \frac{1}{8} & \frac{3}{8} \\
 -\frac{1}{12} & \frac{1}{24} & -\frac{29}{24} \\
\end{array}
\right)$$
and curved \JL{} plural \bkg
\begin{align*}
&\hat\cf=\left(
\begin{array}{ccc}
 \frac{e^{x_3}}{8 \left(e^{\frac{1}{2}
   \left(x_2+x_3\right)}+2\right)} & -\frac{e^{\frac{1}{2}
   \left(x_2+x_3\right)}+1}{8 e^{x_2}+16 e^{\frac{1}{2}
   \left(x_2-x_3\right)}}\\
 \frac{3 \left(e^{\frac{1}{2}
   \left(x_2+x_3\right)}+1\right)}{8 e^{x_2}+16
   e^{\frac{1}{2} \left(x_2-x_3\right)}} & \frac{e^{-x_2}
   \left(4 e^{\frac{1}{2} \left(x_2+x_3\right)}+2
   e^{x_2+x_3}-3\right)}{8 \left(e^{\frac{1}{2}
   \left(x_2+x_3\right)}+2\right)} \\
 \frac{e^{\frac{1}{2} \left(x_2+x_3\right)}
   \left(x_1+x_2+1\right)-3}{8 e^{x_2}+16 e^{\frac{1}{2}
   \left(x_2-x_3\right)}} & \frac{e^{-x_2}
   \left(-e^{\frac{1}{2} \left(x_2+x_3\right)}
   \left(x_1+x_2\right)-e^{x_2+x_3}
   \left(x_1+x_2+2\right)+3\right)}{8 \left(e^{\frac{1}{2}
   \left(x_2+x_3\right)}+2\right)} \\
\end{array}
\right.
\end{align*}
\begin{align}\label{F3vbeta}
&\left.
\begin{array}{ccc}
\frac{e^{\frac{1}{2}
   \left(x_2+x_3\right)} \left(x_1+x_2-3\right)+1}{8
   e^{x_2}+16 e^{\frac{1}{2} \left(x_2-x_3\right)}} \\
 \frac{e^{-x_2}
   \left(e^{x_2+x_3} \left(3 x_1+3
   x_2-2\right)+e^{\frac{1}{2} \left(x_2+x_3\right)}
   \left(3 x_1+3 x_2+8\right)+3\right)}{8
   \left(e^{\frac{1}{2} \left(x_2+x_3\right)}+2\right)} \\
  \frac{e^{-x_2}
   \left(-2 e^{\frac{1}{2} \left(x_2+x_3\right)}
   \left(x_1+x_2+6\right)+e^{x_2+x_3} \left(x_1^2+2
   \left(x_2-1\right) x_1+x_2^2-2
   x_2-14\right)-3\right)}{8 \left(e^{\frac{1}{2}
   \left(x_2+x_3\right)}+2\right)} \\
\end{array}
\right)
\end{align}
with nontrivial torsion. This \bkg\  is obtained from \eqref{JLmtz}-\eqref{piDelta} since from the matrix $M_A{}^B$ we obtain
\begin{equation}
\hat\pi^{ab}=\left(
\begin{array}{ccc}
 0 & 2 e^{-\frac{x_2}{2}-\frac{x_3}{2}}-2 & 2
   e^{-\frac{x_2}{2}-\frac{x_3}{2}}-2 \\
 2-2 e^{-\frac{x_2}{2}-\frac{x_3}{2}} & 0 & 0 \\
 2-2 e^{-\frac{x_2}{2}-\frac{x_3}{2}} & 0 & 0 \\
\end{array}
\right)
\end{equation}
and
$$ \Delta=\frac{1}{4} \left(x_2-x_3\right).$$

Formulas \eqref{hatFA} and \eqref{rhs} with $\tilde{\sigma}=1$ then give
$$  \half \hat F_M  + \begin{pmatrix} \nul_D \\  \frac{\hat f_b{}^{ba} \, \hat v_a^m}{2\tilde{\sigma}} \end{pmatrix}=\left(0,\frac{1}{4},\frac{1}{4},2 e^{-x_3},0,0\right)^T,
$$
so that
\begin{equation}
\hat\varphi(x)=\frac{1}{4}(x_2+x_3),\qquad \hat \cJ=(2 e^{-x_3},0,0).\label{hatphi3v2}
\end{equation}
The \bkg\ \eqref{F3vbeta} together with dilaton 
\begin{equation}
\hat\Phi(x)=-\frac{1}{2} \ln \left(2 e^{-\frac{x_2}{2}-\frac{x_3}{2}}+1\right)
\end{equation}
given by \eqref{eq:dilaton-JL} \emph{satisfy \gsugra s.}

\subsubsection{\JL\ models corresponding to $\{3.x|3\}$ }\label{III.x}

Formulas \eqref{hatFA},\eqref{rhs} for \JLtp y given by the matrix
$$
C_\alpha=\left(
\begin{array}{cccccc}
 0 & 0 & 0 & 1 & -\frac{1}{2} & \frac{1}{2} \\
 \frac{1}{2} & 0 & -1 & 1 & 0 & \frac{1}{2} \\
 -\frac{1}{2} & 0 & 1 & 0 & \frac{1}{2} & 0 \\
 1 & -1 & -1 & -1 & -1 & 0 \\
 0 & 1 & 1 & 1 & 0 & 0 \\
 0 & 1 & 1 & 1 & -1 & 1 \\
\end{array}
\right)
$$
give
$$
\half \hat F_M  + \begin{pmatrix} \nul_D \\  \frac{\hat f_b{}^{ba} \, \hat v_a^m}{2\tilde{\sigma}} \end{pmatrix}=\left(0,0,0,2 e^{x_3-x_2},0,0\right)^T
$$
so that we can choose
$$\hat\varphi(x)=0,\qquad \hat \cJ=(2 e^{x_3-x_2},0,0).$$
The structure of algebra $\{3.x|3\}$ gives us
$$ \hat\pi^{ab}=\left(
\begin{array}{ccc}
 0 & -x_2-x_3 & -x_2-x_3 \\
 x_2+x_3 & 0 & 0 \\
 x_2+x_3 & 0 & 0 \\
\end{array}
\right),\qquad \Delta=\half(x_3-x_2),$$
and we get curved background with torsion
\begin{align*}
\hat\cf=&\frac{2 e^{x_2-x_3}}{x_2^2+2 \left(x_3+1\right)
   x_2+x_3^2+2 x_3-15}  \left(
\begin{array}{ccc}
 1 \\
 \frac{1}{4} \left(4 x_1+x_2+x_3-11\right) \\
 \frac{1}{4} \left(-4 x_1-3 x_2-3 x_3+13\right) \\
\end{array}
\right.
\end{align*}
\begin{align*}
\begin{array}{ccc}
 \frac{1}{4} \left(4 x_1+3 x_2+3 x_3+23\right) \\ \frac{1}{2}
   \left(2 x_1^2+2 \left(x_2+x_3+3\right)
   x_1-x_2^2-x_3^2-4 x_3-2 x_2
   \left(x_3+2\right)-11\right) \\
   \frac{1}{4} \left(-4 x_1^2-2 \left(3 x_2+3
   x_3+5\right) x_1+2 x_2^2+2 x_3^2+x_2+4 x_2
   x_3+x_3+11\right) \\
\end{array}
\end{align*}
\begin{align*}\left.
\begin{array}{ccc}
   \frac{1}{4} \left(-4 x_1-x_2-x_3-17\right) \\  \frac{1}{4} \left(-4
   x_1^2-2 \left(x_2+x_3+3\right) x_1+2 x_2^2+2 x_3^2+3
   x_3+x_2 \left(4 x_3+3\right)+13\right) \\ \frac{1}{2} \left(2 x_1^2+2
   \left(x_2+x_3+1\right) x_1-x_2^2-x_3^2-2 x_2
   \left(x_3-1\right)+2 x_3-7\right) \\
\end{array}
\right)
\end{align*}
that together with
\begin{equation}\label{hatJphi3.x}
\hat\Phi(x)=-\frac{1}{2} \ln\left(x_2^2+2 \left(x_3+1\right) x_2+x_3^2+2 x_3-15 \right),\quad \hat \cJ=(2 e^{x_3-x_2},0,0)
\end{equation}
\emph{satisfy \gsugra s.}

Using the \mtr x
\begin{equation}
\label{Cbeta3x}
C_\beta=\left(
\begin{array}{cccccc}
 0 & 0 & 0 & -1 & 0 & 0 \\
 0 & 0 & -1 & 0 & 0 & 0 \\
 0 & -1 & 0 & 0 & 0 & 0 \\
 -1 & 0 & 0 & 0 & 0 & 0 \\
 0 & 0 & 0 & 0 & 0 & -1 \\
 0 & 0 & 0 & 0 & -1 & 0 \\
\end{array}
\right)
\end{equation}
we get from \eqref{hatFA},\eqref{rhs} 
$$  \half \hat F_M  + \begin{pmatrix} \nul_D \\  \frac{\hat f_b{}^{ba} \, \hat v_a^m}{2\tilde{\sigma}} \end{pmatrix}=\left(0,-\half,\half,0,0,0\right)^T
$$
so that
$$\hat\varphi(x)=\frac{1}{2}(x_3-x_2),\qquad \hat \cJ=(0,0,0).$$
\JL{} \tfn\ of the model \eqref{flatmtz1} then yields torsionless curved background
\begin{align*}
\hat\cf=&\frac{e^{x_2-x_3}}{2\left(x_2^2+2 x_3 x_2+x_3^2-4\right)} \times\\& \left(
\begin{array}{ccc}
 1 & x_1+x_2+x_3 & -x_1+x_2+x_3 \\
 x_1-x_2-x_3 & x_1^2+x_2^2+x_3^2+2 x_2 x_3-8 & -\left(-x_1+x_2+x_3\right){}^2 \\
 -x_1-x_2-x_3 & -\left(x_1+x_2+x_3\right){}^2 & x_1^2+x_2^2+x_3^2+2 x_2 x_3-8 \\
\end{array}
\right)
\end{align*}that together with dilaton
\begin{equation}\label{phi3.x}
\hat\Phi(x)=\frac{1}{2} \ln \left(\frac{ e^{ x_3- x_2}}{x_2^2+2 x_3 x_2+x_3^2-4}\right)
\end{equation} \emph{satisfies \usugra s.}

\subsection{Models obtained by \JLtp y of the  \bkg\ \eqref{flatmtz2}}

Below we present examples obtained by \JL\ \tfn\  of the flat model \eqref{flatmtz2} given by $\{3|1\}$ and 
\begin{equation}
E_0=\left(
\begin{array}{ccc}
 2 & 0 & 1 \\
 0 & -\frac{3}{8} & -\frac{1}{8} \\
 1 & -\frac{1}{8} & \frac{5}{8} \\
\end{array}
\right).
\end{equation} 
It turns out that nearly all models obtained by \JLtp y of the \bkg\ \eqref{flatmtz2} produce either singular $\hat E_0$ or satisfy \usugra s. Ostensible exception is model corresponding to the Leibniz algebra 
$$
\{3v|3\}=\left(\left(III.v,\half\left(T^2-T^3\right)\right), \left(III,0\right)\right).
$$

\subsubsection{\JL\ models corresponding to $\{3.v|3\}$}\label{3to3vflatmtz2}

For the $C_\beta$ \mtr x \eqref{cbeta3v} \JL{} \tfn\ of  the model \eqref{flatmtz2} yields
\begin{align*}
\hat\cf=\left(
\begin{array}{ccc}
 0 & -\frac{3 e^{x_3}}{4 \left(4 e^{\frac{1}{2}
   \left(x_2+x_3\right)}+3\right)} 
   \\
 \frac{5 e^{x_3}}{4 \left(4 e^{\frac{1}{2}
   \left(x_2+x_3\right)}+5\right)} & \frac{32
   e^{\frac{1}{2} \left(x_2+x_3\right)}+80
   e^{x_2+x_3}+15}{1024 e^{x_2}+480 e^{\frac{1}{2}
   \left(x_2-x_3\right)}+512 e^{\frac{1}{2} \left(3
   x_2+x_3\right)}} \\
 \frac{5 e^{x_3}}{4 \left(4 e^{\frac{1}{2}
   \left(x_2+x_3\right)}+5\right)} & -\frac{48 e^{x_2+x_3}
   \left(2 x_1+2 x_2-1\right)+8 e^{\frac{1}{2}
   \left(x_2+x_3\right)} \left(15 x_1+15
   x_2+4\right)+15}{1024 e^{x_2}+480 e^{\frac{1}{2}
   \left(x_2-x_3\right)}+512 e^{\frac{1}{2} \left(3
   x_2+x_3\right)}} \\
\end{array}
\right.\end{align*}
\begin{align}\label{F3vbetaE062}\left.
\begin{array}{ccc}
 -\frac{3 e^{x_3}}
 {4   \left(4 e^{\frac{1}{2} \left(x_2+x_3\right)}+3\right)} \\ \frac{16 e^{x_2+x_3} \left(10 x_1+10
   x_2+3\right)+8 e^{\frac{1}{2} \left(x_2+x_3\right)}
   \left(15 x_1+15 x_2-4\right)-15}{1024 e^{x_2}+480
   e^{\frac{1}{2} \left(x_2-x_3\right)}+512 e^{\frac{1}{2}
   \left(3 x_2+x_3\right)}}\\ \frac{16 e^{x_2+x_3} \left(4 x_1+4
   x_2+5\right)+32 e^{\frac{1}{2}
   \left(x_2+x_3\right)}+15}{1024 e^{x_2}+480
   e^{\frac{1}{2} \left(x_2-x_3\right)}+512 e^{\frac{1}{2}
   \left(3 x_2+x_3\right)}}  
   \\
\end{array}
\right)
\end{align}
that together with
\begin{equation}\label{hatJphi3v2}
 \hat \cJ=(2 e^{-x_3},0,0),\quad \hat\Phi(x)=\frac{1}{2} \ln \left(\frac{ e^{
   x_2+x_3}}{32 e^{\frac{1}{2}
   \left(x_2+x_3\right)}+16
   e^{x_2+x_3}+15}\right)
\end{equation}
or
$$ \hat X =\mathrm{d} \hat \Phi+\hat \cJ\cdot\hat \cF=\frac{2   e^{\frac{1}{2} \left(x_2+x_3\right)}}{32 e^{\frac{1}{2} \left(x_2+x_3\right)}+16 e^{x_2+x_3}+15}\ \mathrm{d}(x_2+x_3)$$
satisfy \gsugra s.

However, in this case we can use gauge \tfn\ \eqref{gauge tfn lambda} to get rid of the Killing vector $\hat \cJ$ because $\mathrm{d}\hat X=0$, and the above given \bkg\ together with dilaton
\begin{equation}\label{arct}
\tilde\Phi= -\arctanh\left(4 \left(e^{\frac{1}{2} \left(x_2+x_3\right)}+1\right)\right),\qquad \mathrm{d}\tilde\Phi=\hat X
\end{equation}
satisfy \usugra s, i.e. \vbe. Analogous situation occurs for the \JL\ \tfn\ given by the \mtr x $C_\gamma$.

Beside that, the \bkg\ \eqref{F3vbetaE062} is torsionless and flat. Vanishing dilaton  and \eqref{arct} are related by the so called $\chi$-symmetry \cite{LHsymm}
\begin{equation}
\label{ambig X} \hat X'_\mu:=\hat X_\mu +\chi_\mu,
\end{equation}
where $\chi = -\hat X$ satisfies
\begin{equation}
\label{cond for chi} \nabla_\nu\chi_\mu=0, \quad (\hat X_\mu +2\,\chi_\mu)\chi^\mu=0.
\end{equation} Other type of reduction to \usugra s was presented in \cite{Saka1803.} where $\cJ\neq 0$ but $\cJ\cdot\cF=0.$
\subsubsection{\JL\ models corresponding to $\{3.x|3\}$ }
\JL{} \tfn\ of the model \eqref{flatmtz2} given by \eqref{Cbeta3x} yields torsionless curved background
\begin{align*}&\hat\cf=\e^{x_2-x_3} \times\\& \left(
\begin{array}{ccc}
 -\frac{1}{2 \left(x_2+x_3-1\right)} & -\frac{2 x_1+x_2+x_3-2}{4 \left(x_2+x_3-1\right)} & \frac{2 x_1+x_2+x_3}{4 \left(x_2+x_3-1\right)} \\
 \frac{-2 x_1+x_2+x_3-2}{4 \left(x_2+x_3-1\right)} & \frac{-4 x_1^2+x_2^2+x_3^2+x_2+2 x_2 x_3+x_3-1}{8 \left(x_2+x_3-1\right)} & \frac{4
   x_1^2+4 x_1-x_2^2-x_3^2+x_2-2 x_2 x_3+x_3+1}{8 \left(x_2+x_3-1\right)} \\
 -\frac{-2 x_1+x_2+x_3}{4 \left(x_2+x_3-1\right)} & \frac{4 x_1^2-4 x_1-x_2^2-x_3^2+x_2-2 x_2 x_3+x_3+1}{8 \left(x_2+x_3-1\right)} & \frac{-4
   x_1^2+x_2^2+x_3^2-3 x_3+x_2 \left(2 x_3-3\right)+3}{8 \left(x_2+x_3-1\right)} \\
\end{array}
\right).
\end{align*}
This \bkg\ satisfies \usugra s  with the dilaton
$$
\hat\Phi(x)=\frac{1}{2} \ln \left(\frac{ e^{ x_3- x_2}}{x_2+ x_3 -1}\right)
$$
obtained from \eqref{eq:dilaton-JL}.

\begin{table}
\begin{center}
\begin{tabular}{|c | c || c | c |}
\hline
{\footnotesize Plural algebra  }& {\footnotesize C-matrix}
&{\footnotesize metric form \eqref{flatmtz1}}&{\footnotesize  metric form \eqref{flatmtz2}}\\
\hline
\hline
{\footnotesize$\{3.v|3\}$}&{\footnotesize $C_\alpha$}&{\footnotesize $\hat E_0$ does not exist}&{\footnotesize $\beta$-\eqn s }\\
{\footnotesize $\{3.v|3\}$}&{\footnotesize $C_\beta$}&{\footnotesize GSUGRA}&{\footnotesize GSUGRA$\rightarrow \beta$-\eqn s }\\
{\footnotesize $\{3.v|3\}$}&{\footnotesize $C_\gamma$}&{\footnotesize $\hat E_0$ does not exist}&{\footnotesize GSUGRA$\rightarrow \beta$-\eqn s }\\
\hline
{\footnotesize $\{3.x|3\}$}&{\footnotesize $C_\alpha$}&{\footnotesize GSUGRA}&{\footnotesize $\hat E_0$ does not exist} \\
{\footnotesize $\{3.x|3\}$}&{\footnotesize $C_\beta$}&{\footnotesize $\beta$-\eqn s}&{\footnotesize $\beta$-\eqn s }\\
\hline
{\footnotesize $\{4.iv|3\}$}&{\footnotesize $C_\alpha$}&{\footnotesize GSUGRA}&{\footnotesize $\hat E_0$ does not exist} \\
{\footnotesize $\{4.iv|3\}$}&{\footnotesize $C_\beta$}&{\footnotesize $\beta$-\eqn s}&{\footnotesize $\beta$-\eqn s }\\
\hline
{\footnotesize $\{5.iii|3\}$}&{\footnotesize $C_\alpha$}&{\footnotesize GSUGRA}&{\footnotesize $\hat E_0$ does not exist} \\
{\footnotesize $\{5.iii|3\}$}&{\footnotesize $C_\beta$}&{\footnotesize $\beta$-\eqn s}&{\footnotesize $\hat E_0$ does not exist}\\
\hline
{\footnotesize $\{6_0.iv|3\}$}&{\footnotesize $C_\alpha$}&{\footnotesize $\beta$-\eqn s}&{\footnotesize $\beta$-\eqn s} \\
{\footnotesize $\{6_0.iv|3\}$}&{\footnotesize $C_\beta$}&{\footnotesize GSUGRA}&{\footnotesize $\hat E_0$ does not exist}\\
{\footnotesize $\{6_0.iv|3\}$}&{\footnotesize $C_\gamma$}&{\footnotesize GSUGRA}&{\footnotesize GSUGRA$\rightarrow \beta$-\eqn s }\\
\hline
{\footnotesize $\{6_{a\pm}.iv|3\}$}&{\footnotesize $C_\beta$}&{\footnotesize $\beta$-\eqn s}&{\footnotesize $\beta$-\eqn s }\\
\hline
\end{tabular}
\caption{Results of \JLtp y of $\{3|1\}$ to type 2 algebras  without spectators. GSUGRA means that models satisfy \gsugra s. $\beta$-\eqn s means that models satisfy \usugra s, i.e. \vbe. We omit results of plurality to $\{6_{a\pm}|3\}$ given by $C_{\alpha}$, since for \eqref{flatmtz1} we were not able to check SUGRA \eqn s in reasonable computer time while for \eqref{flatmtz2} $\hat E_0$ does not exist.}
\label{tab:results}
\end{center}
\end{table}


\section{\JL\ models and \JLtp y with spectators}\label{sec:spec}

Extension of construction of \JL\ models and \JLtp y by spectators can be done similarly as in the case of \pltp y. Namely, first we transform the spectator dependent matrix $E_0(y), y=(y^1,\ldots,y^n)$ by formula
\begin{equation}\label{E0spec}
\wh E_0(y)=\left(\left(\mathcal{P}+ E_0(y) \cdot \mathcal{R}\right)^{-1} \cdot \left(\mathcal{Q}+E_0(y) \cdot \mathcal{S}\right)\right)^T
\end{equation}
where the matrices $\mathcal P,\mathcal Q,\mathcal R,\mathcal S $ are obtained by extension  of the $D\times D$ matrices $P,Q,R,S$ to $(n+D)\times (n+D)$ matrices
\begin{equation}
\nonumber
\mathcal{P} =\begin{pmatrix}\unit_n &0 \\ 0&P \end{pmatrix}, \quad \mathcal{Q} =\begin{pmatrix}\nul_n&0 \\ 0&Q \end{pmatrix}, \quad \mathcal{R} =\begin{pmatrix}\nul_n&0 \\ 0&R \end{pmatrix}, \quad \mathcal{S} =\begin{pmatrix}\unit_n &0 \\ 0& S \end{pmatrix}
\end{equation}
to accommodate the spectator fields. Then we apply formulas for construction of \JL\ models from Section \ref{sec:JLmodels}. Meantime we have to extend the formula for dilaton to 
\begin{align}\nonumber
 \Exp{-2\Phi(x,y)} =& \Exp{-2\hat d(y)} \sqrt{\abs{\det \hat{g}_{ab}(y)}}\Exp{-2\varphi(x)} \Exp{(D-1)\,\Delta(x)} \tilde{\sigma}^{\frac{D-3}{2}}\\ & \abs{\det \left(\left(E_0{}(y)\-1\right){}^{ab}+\pi^{ab}(x)\right) \det\left(a_a{}^b(x)\right)}
\label{eq:specdilaton-JL1}
\end{align}
where $\hat{g}_{ab}(y)$ is symmetric part of $\hat E_0(y)$. On the other hand, at least for $Z^a=0$, it turns out that Supergravity \eqn s are satisfied for
$$
\sqrt{\abs{\det \hat{g}_{ab}(y)}}=\Exp{2\hat d(y)}$$
so that formula for dilaton reads
\begin{align}
\Exp{-2\Phi(x,y)} = \Exp{-2\varphi(x)+(D-1)\,\Delta(x)} \abs{\det \left(\left(E_0{}(y)\-1\right){}^{ab}+\pi^{ab}(x)\right) \det\left(a_a{}^b(x)\right)}.
\label{eq:specdilaton-JL}
\end{align}

\subsection{Four-dimensional flat models corresponding to the algebra $\{3|1\}$}\label{spect}

As mentioned in Section \ref{III.I}, good strategy  for getting models satisfying \gsugra s is finding models corresponding to the algebra of the type 1 with $f_b{}^{ba}=Z^a=0$ and then apply the \JLtp y to the algebras of type 2. Similarly as in Section \ref{III.I}, we will start with flat torsionless model and vanishing dilaton given by the algebra $\{3|1\}={\left(\left({III},-T^2+T^3\right), \left({I},0\right)\right)}$ and some $E_0(t)$\footnote{We have $ n=1$ and $y^1=t$.}.

After rather tedious calculations we were able to find $E_0(t)$ for flat \JL\ model in four dimensions as
$$E_0(t)= \left(
\begin{array}{cccc}
 1 & 2 t & 0 & 0 \\
 2 t & 4 t^2-4 & 0 & 0 \\
 0 & 0 & \frac{1}{2} & 0 \\
 0 & 0 & 0 & \frac{1}{2} \\
\end{array}
\right).$$ \JL\ model then is
\begin{equation}\label{mtzinitx}
\cf=e^{x_2-x_3}\left(
\begin{array}{cccc}
 1 & 2 t & 0 & 0 \\
 2 t & 4 \left(t^2-1\right) & 0 & 0 \\
 0 & 0 & \frac{1}{4} \left(1+ e^{-4 x_1}\right) & -\frac{1}{4}\left(1- e^{-4 x_1}\right) \\
 0 & 0 & -\frac{1}{4}\left(1- e^{-4 x_1}\right) & \frac{1}{4}\left(1+ e^{-4 x_1}\right) \\
\end{array}
\right)\end{equation}
and  $\varphi(x)=x_1+\half(x_3-x_2)$.
 
\subsubsection{\JL\ models corresponding to $\{3.v|3\}$ and spectator $t$}

Formula \eqref{E0spec} for the matrix 
$$C_{\alpha}=\left(
\begin{array}{cccccc}
 0 & 0 & 0 & 1 & -\frac{1}{2} & -\frac{1}{2} \\
 -\frac{1}{4} & -\frac{1}{2} & 0 & 1 & -\frac{1}{2} & -\frac{1}{2} \\
 -\frac{1}{4} & 0 & -\frac{1}{2} & 0 & 0 & 0 \\
 -1 & -2 & -2 & 4 & 0 & -2 \\
 0 & 0 & 0 & -4 & 0 & 2 \\
 0 & 0 & 0 & -4 & 2 & 0 \\
\end{array}
\right)$$ gives
$$\hat E_0(t)= \left(
\begin{array}{cccc}
 0 & -\frac{1}{2 t} & -\frac{3}{4 t} & -\frac{1}{4 t} \\
 \frac{1}{2 t} & \frac{1}{4 t^2} & \frac{3}{8 t^2}-\frac{1}{4} & \frac{1}{8} \left(\frac{1}{t^2}-2\right) \\
 \frac{1}{4 t} & \frac{1}{8} \left(\frac{1}{t^2}+2\right) & \frac{3}{16 t^2}+\frac{1}{8} & \frac{1}{16} \left(\frac{1}{t^2}-4\right)
   \\
 -\frac{1}{4 t} & \frac{1}{4}-\frac{1}{8 t^2} & \frac{1}{4}-\frac{3}{16 t^2} & \frac{1}{8}-\frac{1}{16 t^2} \\
\end{array}
\right).$$
From the algebra $\{3.v|3\}$ we get
\begin{equation}\label{specPoiss3v}
\hat\pi^{ab}=\left(
\begin{array}{cccc}
 0 & 0 & 0 & 0 \\
 0 & 0 & 2
   e^{-\frac{x_2}{2}-\frac{x_3}{2}}-2 &
   2 e^{-\frac{x_2}{2}-\frac{x_3}{2}}-2
   \\
 0 & 2-2
   e^{-\frac{x_2}{2}-\frac{x_3}{2}} & 0
   & 0 \\
 0 & 2-2
   e^{-\frac{x_2}{2}-\frac{x_3}{2}} & 0
   & 0 \\
\end{array}
\right)
\end{equation}
and very complicated torsionless \bkg\ that together with dilaton
$$
\hat \Phi(x,t)= \frac{1}{4}   \log \left(-\frac{9\, t^4\, e^{4 d(t)+x_2-x_3}}{16 \left(4 t^2+2 e^{\frac{1}{2} (x_2+x_3)}+e^{x_2+x_3}-3\right)^2}\right)
$$
satisfy \usugra s for $d(t)= - \ln t + const$.

For the matrix \eqref{cbeta3v} we get  
$$\hat E_0(t)=\left(
\begin{array}{cccc}
 \frac{15}{t^2+15} & \frac{t}{2 t^2+30} & -\frac{3 t}{t^2+15} & \frac{11 t}{t^2+15} \\
 -\frac{t}{2 t^2+30} & \frac{1}{4 t^2+60} & -\frac{3}{2 t^2+30} & -\frac{t^2+4}{2 t^2+30} \\
 -\frac{5 t}{t^2+15} & \frac{5}{2 t^2+30} & \frac{17 t^2+15}{16 t^2+240} & \frac{25-57 t^2}{16
   t^2+240} \\
 \frac{13 t}{t^2+15} & \frac{t^2+2}{2 t^2+30} & \frac{9-41 t^2}{16 t^2+240} & \frac{145
   t^2-113}{16 \left(t^2+15\right)} \\
\end{array}
\right)$$
that together with \eqref{specPoiss3v} give complicated curved \bkg\ with torsion that together with dilaton
$$
\hat \Phi(x,t)=\frac{1}{2} \log \left(\frac{ e^{ x_2+x_3}}{t^2+8 e^{\frac{1}{2}   \left(x_2+x_3\right)}+4 e^{x_2+x_3}+3}\right)
$$
and Killing
$$
\hat\cJ = (0, 2 e^{-x_3}, 0, 0)
$$
satisfy \gsugra s. We get similar results for \JLtp ies to other algebras, see the Table \ref{tab:spectators}.
   
\begin{table}
\begin{center}
\begin{tabular}{|c | c || c | c |}
\hline
{\footnotesize Plural algebra  }& {\footnotesize C-matrix}
&{\footnotesize flat metric in the form \eqref{mtzinitx}}\\
\hline
\hline
{\footnotesize$\{3.v|3\}$}&{\footnotesize $C_\alpha$}&{\footnotesize$\beta$-\eqn s}\\
{\footnotesize $\{3.v|3\}$}&{\footnotesize $C_\beta$}&{\footnotesize GSUGRA}\\
{\footnotesize $\{3.v|3\}$}&{\footnotesize $C_\gamma$}&{\footnotesize GSUGRA}\\
\hline
{\footnotesize $\{3.x|3\}$}&{\footnotesize $C_\alpha$}&{\footnotesize GSUGRA}\\
{\footnotesize $\{3.x|3\}$}&{\footnotesize $C_\beta$}&{\footnotesize $\beta$-equations}\\
\hline
{\footnotesize $\{4.iv|3\}$}&{\footnotesize $C_\alpha$}&{\footnotesize GSUGRA}\\
{\footnotesize $\{4.iv|3\}$}&{\footnotesize $C_\beta$}&{\footnotesize $\beta$-\eqn s}\\
\hline
{\footnotesize $\{5.iii3\}$}&{\footnotesize $C_\alpha$}&{\footnotesize GSUGRA}\\
{\footnotesize $\{5.iii|3\}$}&{\footnotesize $C_\beta$}&{\footnotesize $\beta$-\eqn s}\\
\hline
{\footnotesize $\{6_0.iv|3\}$}&{\footnotesize $C_\alpha$}&{\footnotesize $\beta$-\eqn s}\\
{\footnotesize $\{6_0.iv|3\}$}&{\footnotesize $C_\beta$}&{\footnotesize GSUGRA}\\
{\footnotesize $\{6_0.iv|3\}$}&{\footnotesize $C_\gamma$}&{\footnotesize GSUGRA}\\
\hline
\end{tabular}
\caption{Results of \JLtp y  of $\{3|1\}$ to type 2 algebras  and one spectator.}
\label{tab:spectators}\end{center}
\end{table}


\section{Conclusions}
We have investigated examples of three- and four-dimensional \JL\ models corresponding to Leibniz algebras, introduced in \cite{melsaka} as type 2, i.e. those with $f_b{}^{ba}\neq 0, Z^a=0$. They are displayed in the Table \ref{tab:algebras}. It turned out that these algebras are mutually related by \JLtp y and, moreover, related to algebra
$$
\{3|1\}:=\left(\left(III,-T^2+T^3\right) , \left(I,0\right)\right)
$$
with $f_b{}^{ba}= 0, Z^a=0$.

We have found Jacobi-Lie models that satisfy Generalized Supergravity Equations that need introduction of nontrivial Killing vector. As a side product we have shown that even for type 2 algebras there are also \JL\ models where standard dilaton given by \eqref{eq:dilaton-JL} is sufficient and no Killing vector is necessary for satisfaction of Supergravity Equations. Results are summarized in Tables \ref{tab:results} and \ref{tab:spectators}.

For construction of these models we have used \JLtp y. We have started with flat \bkg s and vanishing dilatons obtained as \JL\ models corresponding to the type 1 algebra $\{3|1\}$, and transformed them to \JL\ models corresponding to the type 2 algebras. Let us note that results of  \JLtp y depend both on matrices $C$ that define the plurality by \eqref{Cplurality} and  form of the flat \bkg\ given by different $E_0$.

Important criterion for decision whether a transformed \bkg, dilaton and Killing satisfy usual or \gsugra s is (non)vanishing of two-form $\mathrm{d}X$ where
$$X_\mu:=\partial_\mu\Phi+\cJ^\kappa\cf_{\kappa\mu},$$
and dilaton $\Phi$ given by \eqref{eq:dilaton-JL} and Killing $\cJ$ by \eqref{rhs}.

If $\mathrm{d}X=0$, the corresponding model satisfies \vbe, while for $\mathrm{d}X\neq 0$ we get models satisfying \gsugra s with nontrivial Killing field given by \eqref{rhs}. It can happen that even if $\cJ\neq 0$, the form $\hat X$ appearing in the \gsugra s \eqref{betaG}-\eqref{betaPhi}  can be (locally) closed and the gauge \tfn\ \eqref{gauge tfn lambda} can change dilaton and bring components of $\cJ$ to zero, so that in fact the model satisfies \usugra s (see e.g. \ref{3to3vflatmtz2}). 

\section*{Appendix - $C$-\mtr ces transforming $\{3|1\}$ to type 2 algebras }

In general we get much greater set of solutions of \eqn s \eqref{tfnX}, but they can be reduced to those given below by choice of free parameters or by change of basis of subalgebras $\bf g$ and $\bf \hat g$.
\begin{enumerate}
\item $C$-matrices for \tfn\ $\{3|1\}\mapsto \{3.v|3\}$\\

\footnotesize
\begin{itemize}\item $C_{\alpha}=\left(
\begin{array}{cccccc}
 0 & 0 & 0 & 1 & -\frac{1}{2} & -\frac{1}{2} \\
 -\frac{1}{4} & -\frac{1}{2} & 0 & 1 & -\frac{1}{2} & -\frac{1}{2} \\
 -\frac{1}{4} & 0 & -\frac{1}{2} & 0 & 0 & 0 \\
 -1 & -2 & -2 & 4 & 0 & -2 \\
 0 & 0 & 0 & -4 & 0 & 2 \\
 0 & 0 & 0 & -4 & 2 & 0 \\
\end{array}
\right),$
\item $C_\beta=\left(
\begin{array}{cccccc}
 0 & \frac{1}{4} & \frac{1}{4} & -1 & \frac{1}{2} & -\frac{1}{2} \\
 0 & 0 & \frac{1}{2} & 0 & 1 & 0 \\
 \frac{1}{2} & 0 & -\frac{1}{2} & 0 & 1 & 0 \\
 1 & 0 & 0 & 0 & 8 & 0 \\
 0 & -\frac{1}{2} & -\frac{1}{2} & 4 & -3 & 3 \\
 0 & -\frac{1}{2} & -\frac{1}{2} & 4 & -1 & 1 \\
\end{array}
\right),$
\item $C_\gamma=\left(
\begin{array}{cccccc}
 0 & \frac{1}{4} & \frac{1}{4} & 0 & 0 & 0 \\
 0 & 0 & \frac{1}{2} & 0 & 0 & 0 \\
 \frac{1}{2} & 0 & -\frac{1}{2} & 0 & 0 & 0 \\
 1 & 0 & 0 & 0 & 4 & 0 \\
 0 & -\frac{1}{2} & -\frac{1}{2} & 2 & -2 & 2 \\
 0 & -\frac{1}{2} & -\frac{1}{2} & 2 & 0 & 0 \\
\end{array}
\right).$
\end{itemize}
\normalsize

\item $C$-matrices for \tfn\ $\{3|1\}\mapsto \{3.x|3\}$\\

\footnotesize
\begin{itemize}
\item $C_\alpha=\left(
\begin{array}{cccccc}
 0 & 0 & 0 & 1 & -\frac{1}{2} & \frac{1}{2} \\
 \frac{1}{2} & 0 & -1 & 1 & 0 & \frac{1}{2} \\
 -\frac{1}{2} & 0 & 1 & 0 & \frac{1}{2} & 0 \\
 1 & -1 & -1 & -1 & -1 & 0 \\
 0 & 1 & 1 & 1 & 0 & 0 \\
 0 & 1 & 1 & 1 & -1 & 1 \\
\end{array}
\right),$
\item $C_\beta=\left(
\begin{array}{cccccc}
 0 & 0 & 0 & -1 & 0 & 0 \\
 0 & 0 & -1 & 0 & 0 & 0 \\
 0 & -1 & 0 & 0 & 0 & 0 \\
 -1 & 0 & 0 & 0 & 0 & 0 \\
 0 & 0 & 0 & 0 & 0 & -1 \\
 0 & 0 & 0 & 0 & -1 & 0 \\
\end{array}
\right).$
\end{itemize}
\normalsize

\item $C$-matrices for \tfn\ $\{3|1\}\mapsto \{4.iv|3\}$\\

\footnotesize
\begin{itemize}
\item $C_\alpha=\left(
\begin{array}{cccccc}
 0 & 0 & 0 & 1 & 0 & 0 \\
 0 & 0 & 1 & 0 & 0 & 0 \\
 0 & 0 & -1 & 0 & -1 & 0 \\
 1 & 0 & 0 & 0 & 1 & 0 \\
 0 & -1 & -1 & 1 & -1 & 1 \\
 0 & -1 & -1 & 1 & 0 & 0 \\
\end{array}
\right),$
\item $C_\beta=\left(
\begin{array}{cccccc}
 0 & 0 & 0 & -1 & \frac{1}{2} & -\frac{1}{2} \\
 -\frac{1}{2} & -1 & 0 & -1 & \frac{1}{2} & 0 \\
 \frac{1}{2} & 0 & -1 & 0 & -\frac{1}{2} & 0 \\
 -1 & 0 & 0 & 0 & 1 & 0 \\
 0 & 0 & 0 & 0 & -1 & 0 \\
 0 & 0 & 0 & 0 & 0 & -1 \\
\end{array}
\right).$
\end{itemize}
\normalsize

\item $C$-matrices for \tfn\ $\{3|1\}\mapsto \{5.iii|3\}$\\

\footnotesize
\begin{itemize}
\item $C_\alpha=\left(
\begin{array}{cccccc}
 0 & 0 & 0 & 1 & 0 & 0 \\
 0 & -\frac{1}{2} & \frac{1}{2} & \frac{1}{2} & \frac{1}{2} &
   \frac{1}{2} \\
 0 & \frac{1}{2} & -\frac{1}{2} & -\frac{1}{2} & 0 & 0 \\
 1 & 0 & 0 & 0 & 1 & 0 \\
 0 & 1 & 1 & -1 & 0 & 0 \\
 0 & 1 & 1 & -1 & 1 & -1 \\
\end{array}
\right),$
\item $C_\beta=\left(
\begin{array}{cccccc}
 0 & 0 & 0 & -1 & \frac{1}{2} & -\frac{1}{2} \\
 -\frac{1}{2} & -1 & 0 & 0 & 0 & 0 \\
 \frac{1}{2} & 0 & -1 & 0 & 0 & 0 \\
 -1 & 0 & 0 & 0 & 0 & 0 \\
 0 & 0 & 0 & 0 & -1 & 0 \\
 0 & 0 & 0 & 0 & 0 & -1 \\
\end{array}
\right).$
\end{itemize}
\normalsize

\item $C$-matrices for \tfn\ $\{3|1\}\mapsto \{6_0.iv|3\}$\\

\footnotesize
\begin{itemize}
\item $C_\alpha=\left(
\begin{array}{cccccc}
 0 & 0 & 0 & -1 & -\frac{1}{2} & \frac{1}{2} \\
 \frac{1}{2} & -1 & 0 & 0 & 0 & 0 \\
 -\frac{1}{2} & 0 & -1 & 0 & 0 & 0 \\
 -1 & 0 & 0 & 0 & 0 & 0 \\
 0 & 0 & 0 & 0 & -1 & 0 \\
 0 & 0 & 0 & 0 & 0 & -1 \\
\end{array}
\right),$
\item $C_\beta=\left(
\begin{array}{cccccc}
 0 & 0 & 0 & 1 & -1 & 1 \\
 -1 & 0 & 1 & \frac{1}{2} & -\frac{1}{2} & \frac{1}{2} \\
 1 & 0 & -1 & 0 & 1 & 0 \\
 1 & -\frac{1}{2} & -\frac{1}{2} & \frac{1}{2} & 1 & 0 \\
 0 & 1 & 1 & -1 & 0 & 0 \\
 0 & 1 & 1 & -1 & 1 & -1 \\
\end{array}
\right),$
\item $C_\gamma=\left(
\begin{array}{cccccc}
 0 & 0 & 0 & 1 & -1 & 1 \\
 -1 & 0 & 1 & 1 & 1 & 1 \\
 1 & 0 & -1 & 0 & 0 & 0 \\
 1 & -\frac{1}{2} & -\frac{1}{2} & -1 & -1 & -1 \\
 0 & \frac{1}{2} & \frac{1}{2} & 1 & -1 & 1 \\
 0 & \frac{1}{2} & \frac{1}{2} & 1 & 0 & 0 \\
\end{array}
\right).$
\end{itemize}
\normalsize

\item $C$-matrices for \tfn\ $\{3|1\}\mapsto \{6_{a\pm}|3\}$\\

\footnotesize
\begin{itemize}
\item $C_\alpha=\left(
\begin{array}{cccccc}
 0 & 0 & 0 & 1 & \frac{a_\pm-1}{2} & \frac{1-a_\pm}{2}
   \\
 \frac{1-a_\pm}{2} & 0 & -1 & \frac{1}{1-a_\pm} &
   -\frac{a_\pm}{2} & \frac{1}{2} \\
 \frac{a_\pm-1}{2} & 0 & 1 & 0 & \frac{a_\pm+1}{2} & 0
   \\
 1 & \frac{1}{a_\pm-1} & \frac{1}{a_\pm-1} &
   \frac{1}{1-a_\pm} & \frac{a_\pm}{a_\pm-1} &
   \frac{1}{1-a_\pm} \\
 0 & 1 & 1 & -1 & 1 & -1 \\
 0 & 1 & 1 & -1 & 0 & 0 \\
\end{array}
\right),$
\item $C_\beta=\left(
\begin{array}{cccccc}
 0 & 0 & 0 & -1 & \frac{1}{2}{a_\pm} & -\frac{1}{2}{a_\pm} \\
\frac{1}{2}{a_\pm} & 0 & -1 & 0 & 0 & 0 \\
 -\frac{1}{2}{a_\pm} & -1 & 0 & 0 & 0 & 0 \\
 -1 & 0 & 0 & 0 & 0 & 0 \\
 0 & 0 & 0 & 0 & 0 & -1 \\
 0 & 0 & 0 & 0 & -1 & 0 \\
\end{array}
\right).$
\end{itemize}
\end{enumerate}

\end{document}